\documentstyle[a4,amsmath,psfig,12pt]{article}
\begin{document}

\title{Shocks in non-loaded bead chains with impurities}
\author{Erwan Hasco\"et$^{1}$ and Hans J. Herrmann$^{1,2}$}
\date{\today}
\pagestyle{plain}
\maketitle
\centerline{
$^1$P.M.M.H. Ecole Sup\'erieure de Physique et Chimie Industrielles,}
\centerline{10, rue Vauquelin, 75231 Paris CEDEX 05 France}
\centerline{$^2$ICA1, University of Stuttgart, Pfaffenwaldring 27, D-70569
Stuttgart Germany}
\begin{abstract}
We numerically investigate the problem of the propagation of a shock
in an horizontal non-loaded granular chain with a bead interaction
force exponent varying from unity to large values. When $\alpha$ is
close to unity we observed a cross-over between a
nonlinearity-dominated regime 
and a solitonic one, the latest being the final steady state of the
propagating wave. In the case of large values of $\alpha$ the
deformation field given by the numerical simulations is completely
different from the one obtained by analytical calculation. In the
following we studied the interaction of these shock waves with a mass
impurity placed in the bead chain. Two different physical pictures
emerge whether we consider a light or a heavy impurity mass. The
scatter of the shock wave with a light impurity yields damped
oscillations of the impurity which then behave as a solitary wave
source. Differently an heavy impurity is just shifted by the shock and
the transmitted wave loses its solitonic character being fragmented
into waves of decreasing amplitudes.
\end{abstract}

\section{Introduction}
\label{intro}

Since the pioneering work of Nesterenko \cite{Nest1} there has been an
increase of activity devoted to the study of wave propagation in
granular media. The theoretical prediction of solitons-like waves
propagating in one dimensional chains of perfect elastic spheres has
been experimentally checked by Lazaridi and Nesterenko \cite{NestLaz}
and more recently by Coste, Falcon and Fauve \cite{CFF}. These studies
concern the idealized case of spheres interacting according to the
Hertz law where the interaction force depends on the overlap distance
of the spheres to the power $3 \over 2$. The result of Nesterenko for
the particular Hertzian case can actually be generalized to an
interaction law where the exponent can vary from one to infinity
\cite{Nest2}, the case of the exponent equal to unity being special
since it leads to a self-similar asymmetric pulse propagation resulting
from a shock perturbation \cite{HHL,HS}. Instead of traveling as a
soliton the pulse propagates increasing its width and decreasing its
amplitude. It was then showed that the amplitude decreases like
$t^{-0.17}$ and the width increases like $t^{1 \over 3}$. This
generalization of the interaction law by letting the interaction
exponent vary allows us to take into account the various shapes of two
interacting beads. 

Besides isolating non-linearities in one dimensional chains other
physical ingredients have been added in order to make these models
more realistic. Sinkovits and Sen \cite{Sin} did molecular dynamics
simulations on gravitationally compacted Hertzian chains. It was then observed
that gravitation makes waves loose their soliton-like features
becoming solitary waves with the wave speed increasing
with depth $z$ as $z^{1\over6}$ as predicted by Hertz law. Besides, if
an impurity is placed in the chain, the propagating wave is scattered
by it and a reflected wave is obtained in addition to the transmitted
wave, these waves having a similar shape as the incident
wave. Following these first results Hong, Ji and Kim \cite{HJK} extracted
several power laws for the velocity amplitude of a wave propagating in
a gravitationally compacted chain. It was shown that in a homogeneous
chain of beads the amplitude velocity of the solitary pulse decreases
like $z^{-{1 \over 4}({1 \over 3} + {1 \over \alpha})}$ for an interaction
law exponent $\alpha$ greater than unity. This power law is only valid
in the case of a small perturbation where the non-linear interaction
law can be locally linearized, the stiffness coefficient then varying
with depth.
 
In this paper, we present molecular dynamics results for non-loaded
horizontal granular chains containing impurities. In our model the
force interaction law exponent is allowed to vary from one to
infinity, the Hertzian case being considered as a special one. After
having reviewed the theoretical results for shock propagation in a
horizontal bead chain and their relevance in reproducing the results
from numerical simulations we study the interaction of shock waves
with impurities of various masses. We show that the patterns observed
for the scattered waves vary as impurity masses are larger or smaller
than the mass of the beads constituting the granular chain. In the
case of a light impurity we observed a backscattered solitary wave
followed by an increasing number of secondary waves of decreasing
amplitudes produced at the impurity location. These secondary waves
also propagate in the forward direction following the transmitted
wave. We looked at the displacement of the impurity after the
collision. The impurity actually oscillates in a ballistic motion
around its equilibrium position. These oscillations are damped since
the impurity collides its nearest neighbors producing a backward and a
forward moving pulse in each cycle. If we consider the interaction of
a shock with an heavy impurity we obtain a very different
scenario. Instead of exciting the impurity the shock only shifts it in
its moving direction. The chain is then halved in two separated chains
on which propagate several pulses. The situation is no more symmetric
as in the previous case. After the collision the incident wave does
not produce a stable transmitted wave but is decomposed in a series of
pulses of decreasing amplitudes. Before the impurity we observe the
propagation of a stable reflected solitary wave.  In each case the
amplitude of the reflected pulse grows with the absolute value of the
difference between the mass of the impurity and the normal mass. Hence
the speed of the reflected pulse grows with this mass difference since
the speed of our solitary waves increases with the amplitude.

In section $2$ we present the model and its analytical solution
in the continuum limit. This solution is compared to the numerical simulations
as the interaction law exponent is tuned from unity to large values.
In section $3$ we analyze the interaction of these shock waves with an
impurity and display the energy distribution in the bead chain after
the scattering with the impurity. A conclusion is given in section $4$.

\section{Theoretical results versus numerical simulations}
\label{sec:1}
The horizontal granular chain consists of $N$ touching beads of equal
sizes and masses. The repulsive interaction force between two adjacent
beads under compression is written $F=k{\delta^{\alpha}}$ where $\delta$
is the overlap of the beads. For the exponent
$\alpha$ equal to $3\over2$ we have the Hertz law for perfectly
spherical beads and the stiffness constant $k$ can be expressed as a
function of the spheres radii, the Young modulus and the Poisson ratio of the
material. Here we let $\alpha$ vary from one to infinity considering
beads with different geometries. The equation of motion for the
displacement $u_i$ of the bead $i$ is given by
\begin{equation}
m\ddot{u}_i=k\delta_{i-1,i}^{\alpha}\theta(\delta_{i-1,i})-
k\delta_{i,i+1}^{\alpha}\theta(\delta_{i,i+1}) \quad 0 < i < N-1
\end{equation}
In terms of the displacements of the beads $i-1$ and $i+1$ one can write
\begin{equation}
m\ddot{u}_i=k(u_{i-1}-u_i)^{\alpha}\theta(u_{i-1}-u_i) 
 -k(u_i-u_{i+1})^{\alpha}\theta(u_i-u_{i+1})
\end{equation}
the Heaviside function having been introduced in order to have
no tension forces.

This equation has been solved by Nesterenko \cite{Nest2} in the case
of a shock perturbation by doing a long wavelength development. This
approximation will be valid if the spatial extension of the wave
generated by the shock is much greater than the size of a single bead.
Assuming this we write $u_i$ as $u(x,t)$. Moreover, since we are
solving the equation for a wave in which the beads are compressed one
can get rid of the Heaviside functions. The resulting equation is
written
\begin{multline}
mu_{tt}=k\Big(-au_{x} + {a^2 \over 2} u_{xx} - {a^3 \over 6}
u_{xxx} + {a^4 \over 24} u_{xxxx}\Big)^\alpha \\
-k\Big(-a u_x - {a^2 \over 2} u_{xx} - {a^3  \over 6 } u_{xxx} - {a^4
\over 24} u_{xxxx}\Big)^\alpha
\end{multline}
where a is the dimension of the bead and we developed $u(x+a)$ and
$u(x-a)$ to fourth order.
We then rewrite the equation
\begin{multline}
{m \over k}u_{tt}=(-au_x)^\alpha\bigg[\Big(1-{a \over
2}{u_{xx}\over u_{x}}+{a^2 \over 6}{u_{xxx} \over u_{x}}-{a^3 \over
24}{u_{xxxx}\over u_x}\Big)^\alpha \\
 -\Big(1+{a \over 2}{u_{xx}\over
u_x}+{a^2\over 6}{u_{xxx}\over u_{x}}+{a^3\over 24}{u_{xxxx}\over
u_{x}}\Big)^\alpha\bigg], 
\end{multline}
\begin{equation}
-u_x>0
\label{cond}
\end{equation}
$u_x$ being of order ${u \over L}$ where $L$ is the dimension of the
pulse and expand the powers in Taylor series to the order $1 \over
L^3$. We then obtain
\begin{multline}
{m \over k}u_{tt}=a^\alpha\Big(\alpha a(-u_x)^{\alpha-1}u_{xx}+\alpha
{a^3 \over 12}(-u_x)^{\alpha-1}u_{xxxx} \\
-\alpha(\alpha-1){a^3 \over
6}{u_{xx}u_{xxx} \over
(-u_x)^{2-\alpha}}+\alpha(\alpha-1)(\alpha-2){a^3 \over 24}{{u_{xx}}^3
\over (-u_x)^{3-\alpha}}\Big)
\label{eq}
\end{multline}
Looking for traveling wave solutions $u(\xi) \equiv u(x-Vt)$ with $V$ being the
speed of the wave, a longer calculation \cite{Nest2} gives us for the deformation 
\begin{equation}
-u_x=\Big({\alpha+1 \over 2}{V^2 \over {c_{\alpha}}^2 } \Big)^{1
 \over \alpha-1}\bigg\arrowvert \cos{\Big({\alpha-1 \over
 \sqrt{\alpha(\alpha+1)}}{\sqrt6 \over
 a}\xi\Big)}\bigg\arrowvert^{2 \over \alpha-1}
\label{cos}
\end{equation}
With
\begin{equation}
{c_{\alpha}}^2={a^{\alpha+1}k \over m}
\end{equation}
This periodic
wave solution is singular since it violates the condition $-u_x>0$ in
eq.(\ref{cond}). We actually get the shock wave solution by imposing
the argument of the cosines to vary in the interval $[-{\pi \over
2}(2k+1),{\pi \over 2}(2k+1)]$, $k \in Z$. From the expression above
we easily extract the wave speed $V$:
\begin{equation}
V=\bigg({2{c_{\alpha}}^2 \over \alpha+1}\bigg)^{1 \over \alpha+1}
v_{max}^{\alpha-1 \over \alpha+1}
\label{vf}
\end{equation}
$v_{max}$ being the maximum of the velocity pulse.

\begin{figure}[htbp]
\centerline{
         \psfig{figure=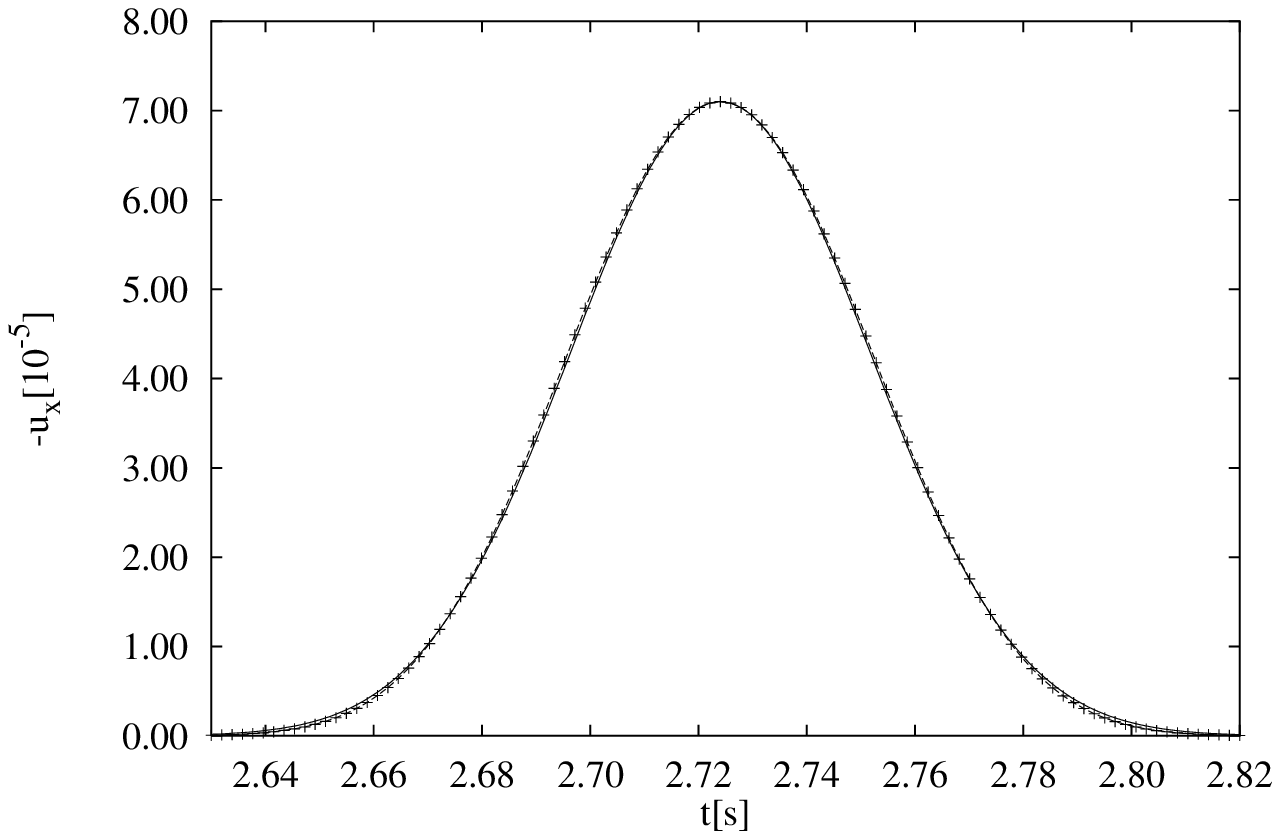,width=\textwidth}}   

\caption[]{Deformation curves of the bead number $100$ in a chain of
$200$ beads, the force exponent being equal to $1.2$. There is very good
agreement between numerical and theoretical results. The crossed curve
from the analytical result eq.(\ref{cos}) superimposes the
plain line curve from the numerical simulation. The crossed curve
obviously correspond to a single oscillation of the periodic solution eq.(\ref{cos}).}
\label{cos1.2}
\end{figure}

\begin{figure}[htbp]
\centerline{
         \psfig{figure=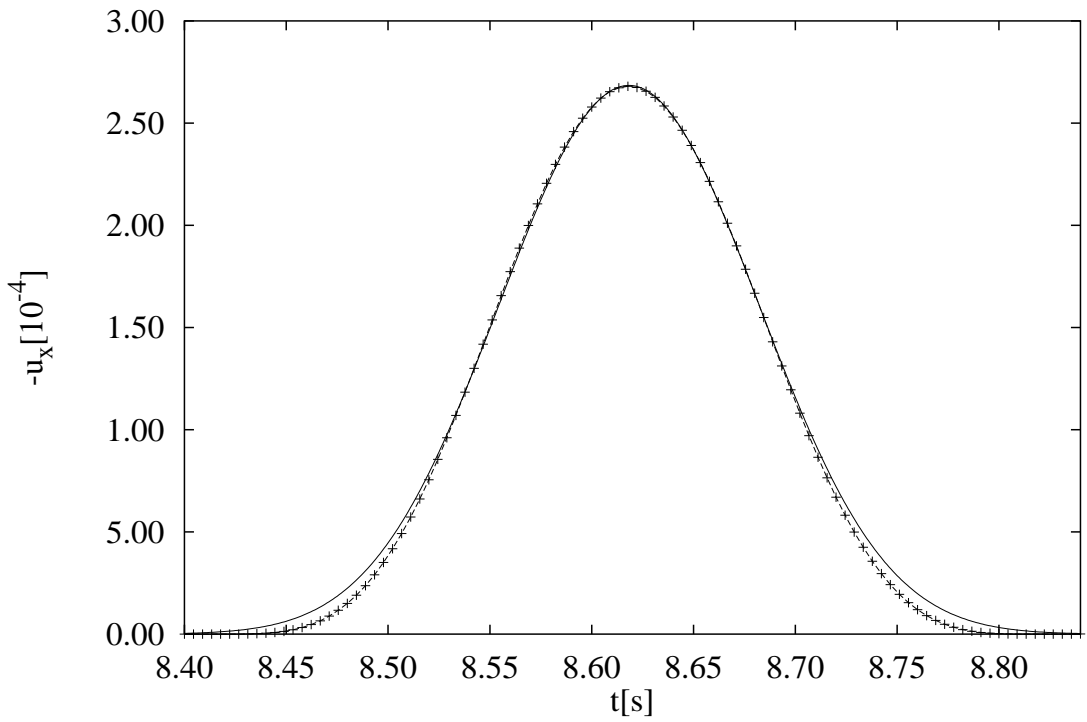,width=\textwidth}}   

\caption[]{Deformation curves when $\alpha=1.5$. Discrepancies between
numerical and theoretical results appear at the curves tails. The
crossed tails of the analytical result remain below the real tails of
the numerical simulation.}
\label{cos1.5}
\end{figure}

\begin{figure}[htbp]
\centerline{
         \psfig{figure=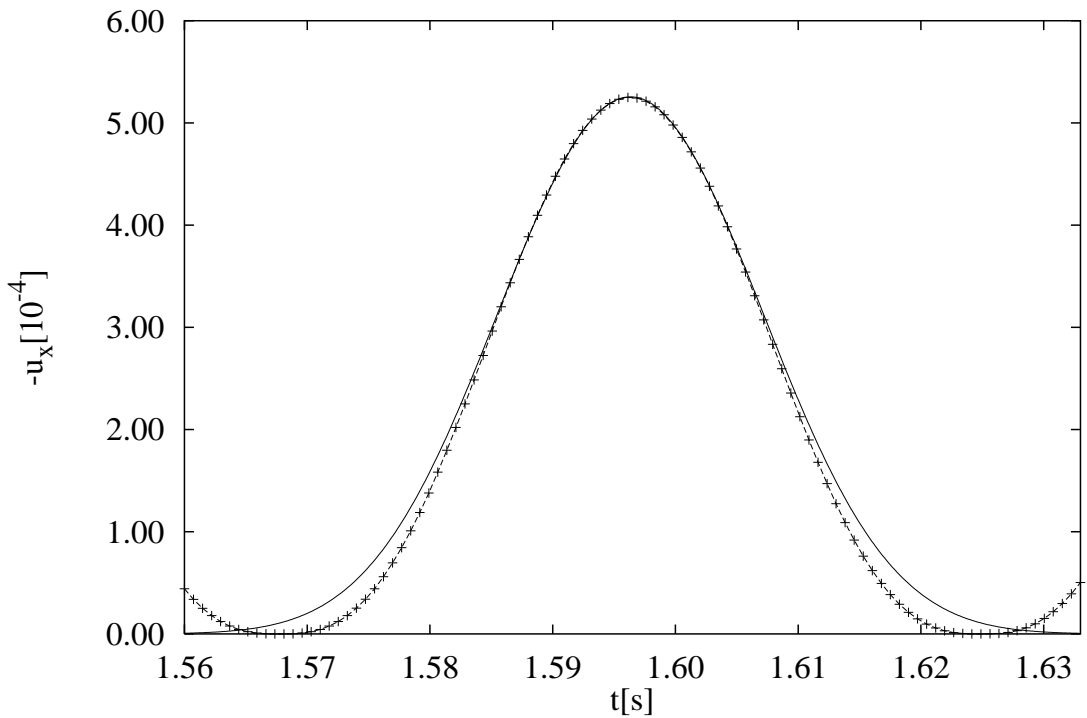,width=\textwidth}}   

\caption[]{Deformation curves when $\alpha=1.7$. The increase in
$\alpha$ increases the deviation of the curves at the tails. The
oscillating character of the analytical result is displayed at the
corners of the plot where one can see the end of the previous
oscillation of the non-linear periodic solution on left and the beginning of its next oscillation on right.  }
\label{cos1.7}
\end{figure}

\begin{figure}[htbp]
\centerline{
         \psfig{figure=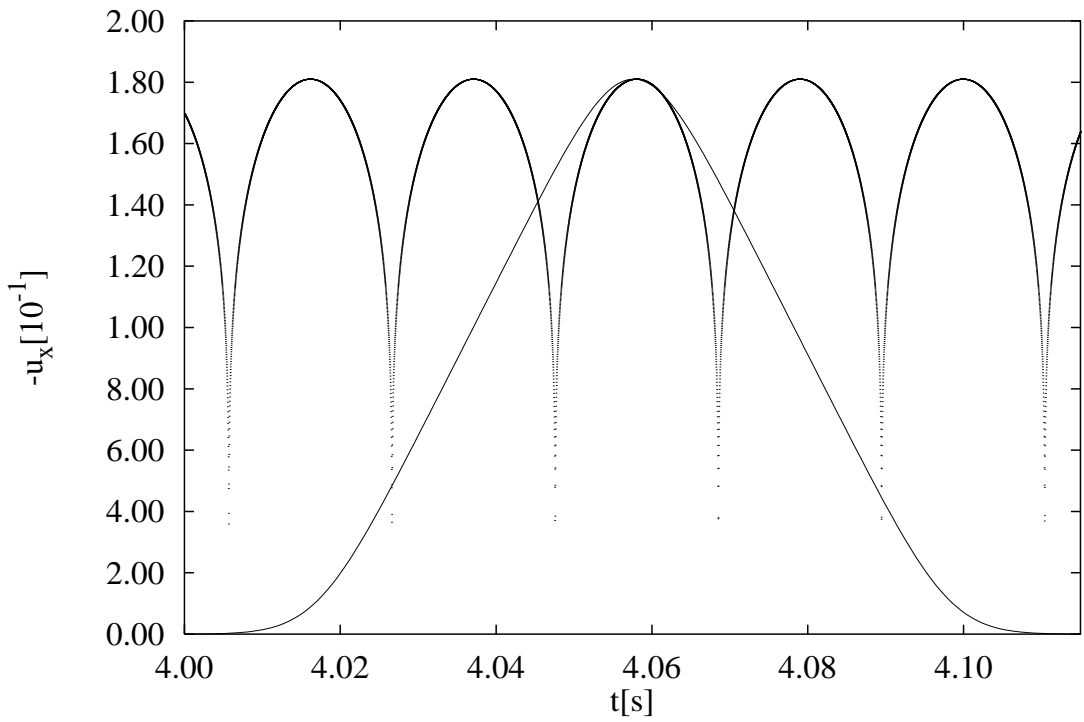,width=\textwidth}}   

\caption[]{Deformation curves when $\alpha=10$. There is no agreement
between analytical and numerical results. The true
deformation has a triangular shape while the analytical curve display
a periodic parabola shape.}

\label{cos10}
\end{figure}

\begin{figure}[htbp]
\centerline{
         \psfig{figure=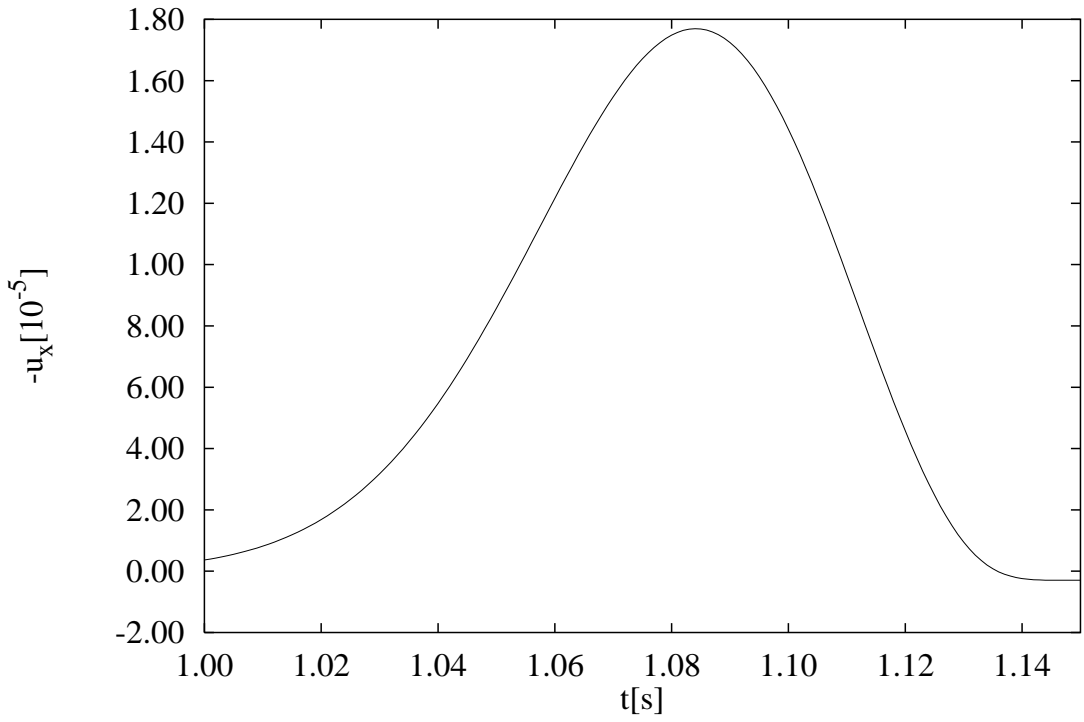,width=\textwidth}}   

\caption[]{Deformation curve when $\alpha=1.01$ for bead number $100$.
The pulse is not symmetric. We are in the non-linearity dominated regime.}
\label{cos1.01}
\end{figure}

\begin{figure}
\centerline{
         \psfig{figure=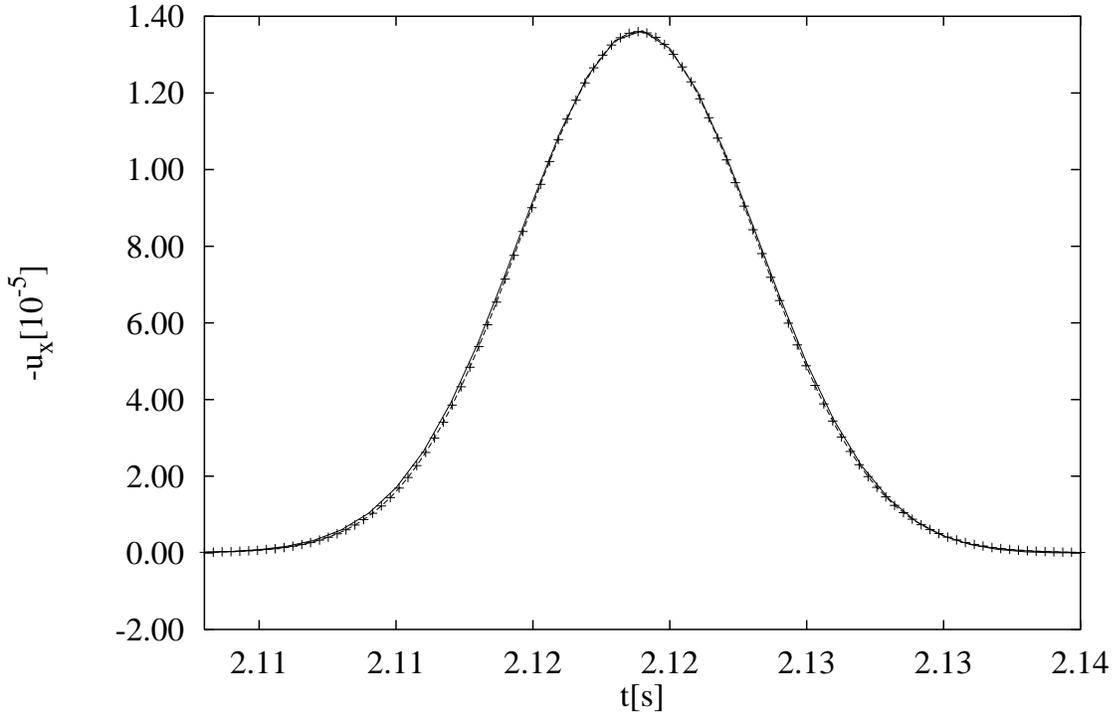,width=\textwidth}}   

\caption[]{Deformation curves when $\alpha=1.01$ for bead number
$2000$ of a bead chain of $3000$ beads. The pulse reached its
stationary soliton shape given by eq.(\ref{cos}) and the numerical
and analytical curves superimpose. }
\label{cos1.01.b}
\end{figure}

\begin{figure}[htbp]
\centerline{
         \psfig{figure=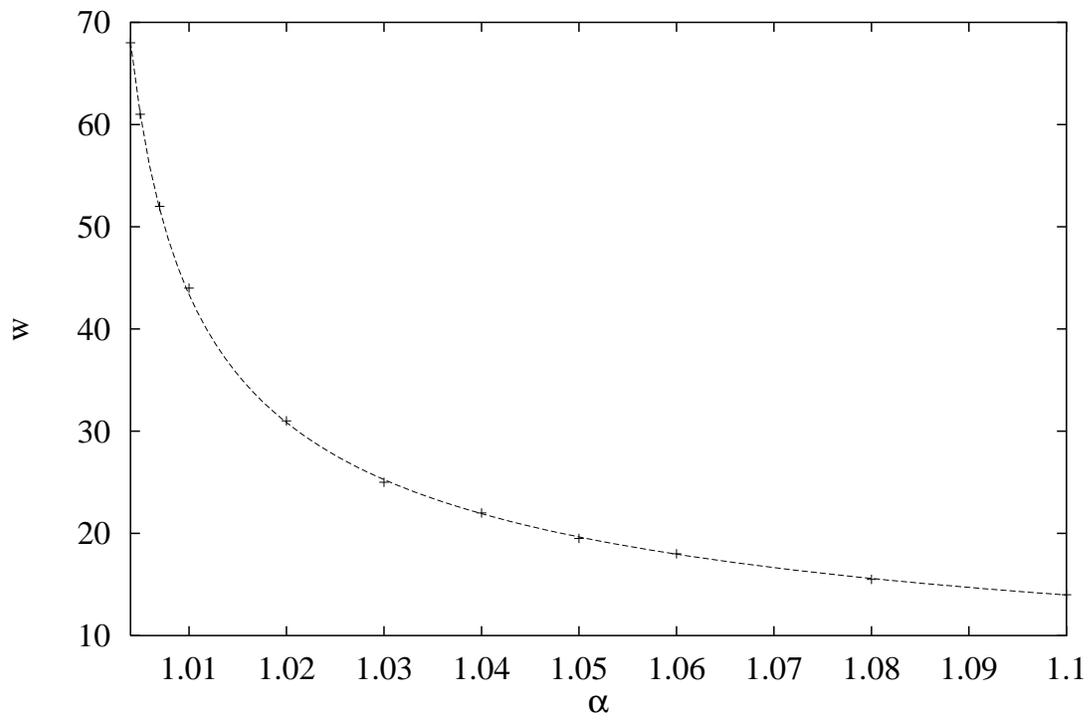,width=\textwidth}}   

\caption[]{Plot of the width of the pulse as function of $\alpha$
when $\alpha$ is close to unity. The dashed line displays the
corresponding power law : $w \propto (\alpha-1)^{-0.492}$.}
\label{width}
\end{figure}

\begin{figure}[htbp]
\centerline{
         \psfig{figure=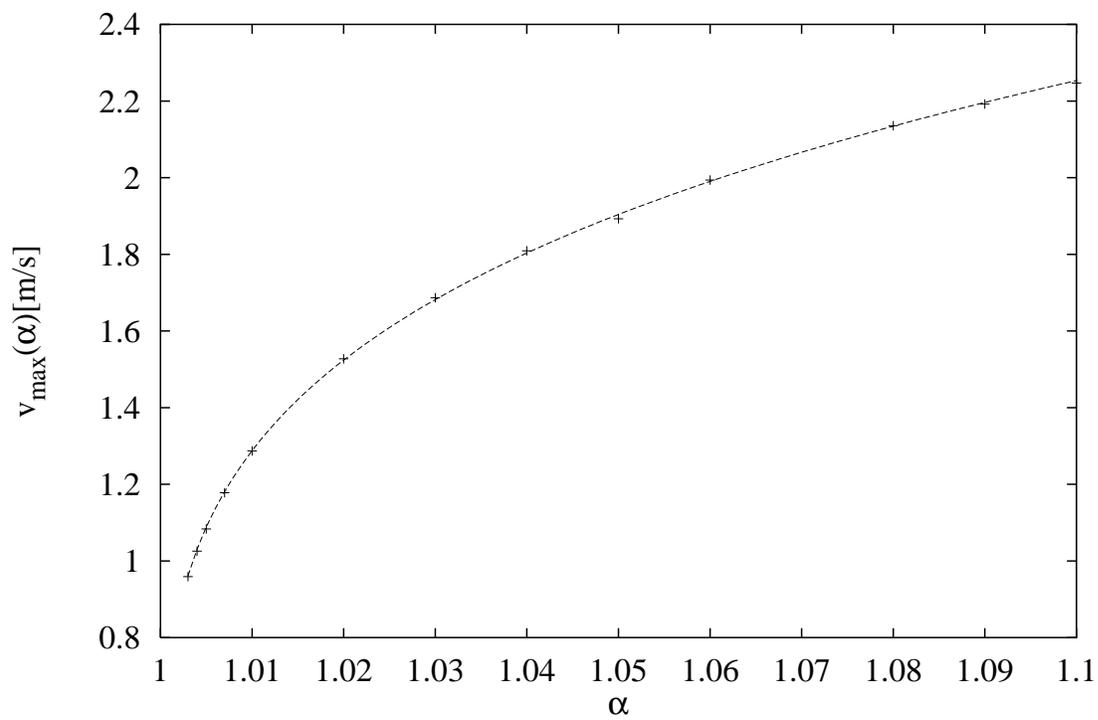,width=\textwidth}}   

\caption[]{Plot of the velocity pulse amplitude as function of $\alpha$ when
$\alpha$ is close to unity. The dashed line displays the corresponding 
power law : $v_{max} \propto (\alpha-1)^{0.243}$.}
\label{vmax}
\end{figure}

The accuracy of this analytical approximation has been tested
numerically with a Gear fifth order predictor-corrector algorithm on a
horizontal chain of $200$ beads with free boundary conditions. A shock
perturbation is imposed on the chain by giving the first bead a $5
m.s^{-1}$ speed at zero time. We choose beads of mass $m=1g$ and
stiffness $k=10^7 N.m^{-1}$. We first looked at the shape of the
deformation profile of the bead number $100$ as function of time and
compared it with with the shape given by the expression (\ref{cos})
for different values of $\alpha$. We displayed in fig.(\ref{cos1.2}),
fig.(\ref{cos1.5}) and fig.(\ref{cos1.7}) the superposition of
numerical and analytical curves for $\alpha$ equal to $1.2$, $1.5$ and
$1.7$. One easily sees on these curves that the analytical results are
rather good for low values of $\alpha$ $(\alpha>1)$ showing
discrepancies when $\alpha$ is increased. Actually, the tails of the
deformations curves are much longer than the ones given by equation
(\ref{cos}). If the value of $\alpha$ is increased to higher values
the numerical and analytical results become completely different as
one can see in fig.(\ref{cos10}) where we took $\alpha$ equal to
$10.0$. Other discrepancies can be found if $\alpha$ is taken in the
neighborhood of unity. It can be seen in fig.(\ref{cos1.01}) that the
deformation curve on the bead number $100$ is not symmetric when
$\alpha$ is taken equal to $1.01$. This non-symmetric shape actually
evolves to a symmetric one as the pulse propagate.  We simulated a
chain of $3000$ beads and considered the deformation of the bead
number $2000$ : the pulse has the soliton shape given by equation
(\ref{cos}) (fig.(\ref{cos1.01.b})). The pulse has actually to
propagate over $2000$ beads before reaching its solitonic regime
predicted by (\ref{cos}). In fact, there is a cross-over between a
nonlinearity-dominated regime and a solitonic regime where dispersion
is balanced by nonlinearity. The non-linearity dominated regime
corresponds to a transient where the amplitude of the pulse decreases
slowly in time before reaching a constant value, i.e, the solitonic
regime. At the same time the width of the pulse increases also
reaching a stationary value. Moreover the shape of the pulse evolves
continuously from a non-symmetric one to a symmetric one. The time of
the transient obviously increases as we decrease $\alpha$ (if we take
$\alpha=1.004$ the pulse has to propagate over $15000$ beads before
reaching its stationary value).  This cross-over behavior can
be understood by looking at equation (\ref{eq}) for $\alpha$ close to
one.  We can rewrite this equation with a small parameter $\epsilon$
where $\alpha=1+\epsilon$ :
\begin{equation}
{m \over k}u_{tt}=a^{2+\epsilon}(-u_x)^{\epsilon}\Big(u_{xx}+ {a^2
\over 12}u_{xxxx}\Big)
\end{equation}
If we consider $(-u_x)^{\epsilon}$ constant equal to one we obtain the
long wavelength
equation for the linear springs chain which is the same as the simple
wave equation up to a dispersive term, the fourth order derivate of
the displacement. It has been shown \cite{HHL,HS}
that the velocity solution of this equation is not a soliton but a
non-symmetric self-similar wave of decreasing amplitude and increasing
width, the amplitude going to zero and the width going to infinity as
time goes to infinity. For $\alpha$ close to one the non-linearity
$(-u_x)^{\epsilon}$ vary slowly and the time for it to balance the
action of the
dispersive four order term increases as $\epsilon$ goes to zero. We can
actually calculate how the amplitude and the width of the velocity pulse
vary with epsilon. The velocity of a bead is given by eq.(\ref{cos})
up to the factor $V$. Replacing $\alpha$ with $\epsilon+1$ we get :
\begin{equation}
v=A\bigg\arrowvert
\cos{\Big({{\sqrt{6}\epsilon}\over{a\sqrt{(\epsilon+1)(\epsilon+2)}}}\xi\bigg)}\Big\arrowvert^{2
\over \epsilon}
\label{speed}
\end{equation}
where
\begin{equation}
A=\Big({\epsilon+2 \over 2}{V^{\epsilon+2} \over c^2_{\epsilon+1}}\Big)^{1
\over \epsilon}
\end{equation}
We then write eq.(\ref{speed}) in an exponential form and expand its
argument in a Taylor series keeping the lower order term :
\begin{equation}
v=A\exp{\Big(-{6\epsilon \over a^2(\epsilon+1)(\epsilon+2)}\xi^2\Big)}
\end{equation}
The solution of eq.(\ref{eq}) is then approximated by a Gaussian for
$\alpha$ close to unity.
According to the properties of Gaussian distributions the width $w$ of
the solitary wave follows the scaling law $w\propto{\epsilon}^{-{1 \over
2}}$, i.e :
\begin{equation}
w\propto(\alpha-1)^{-{1 \over 2}}
\end{equation}
For $\alpha=1$ we recover the result of \cite{HHL,HS} where the
width of the wave was found to diverge as time goes to infinity. We
checked numerically the accuracy of this scaling relation and found a
good agreement (see Fig.(\ref{width})). A second scaling relation can
be obtained by using the conservation of the kinetic energy in the
wave since the beads following the wave have negligible velocities. It
follows that the product $v^2_{max}w$ remains constant. Hence,
\begin{equation}
v_{max}\propto(\alpha-1)^{1 \over 4}
\end{equation}
We again recover the result of \cite{HHL,HS} where the velocity
amplitude was found to go to zero as time goes to infinity. We
displayed in Fig.(\ref{vmax}) the value of the final velocity amplitude
$v_{max}$ when $\alpha$ is varied in the neighborhood of one and found
again a good agreement with the obtained scaling relation. 
These two scaling laws show that the transition between the
self-similar regime $\alpha=1$ to the solitonic regime
$\alpha>1$ is actually continuous, the value $\alpha=1$ being not
singular.
 

\section{Collision of a shock wave with an impurity}

\begin{figure}[htbp]
\centerline{
         \psfig{figure=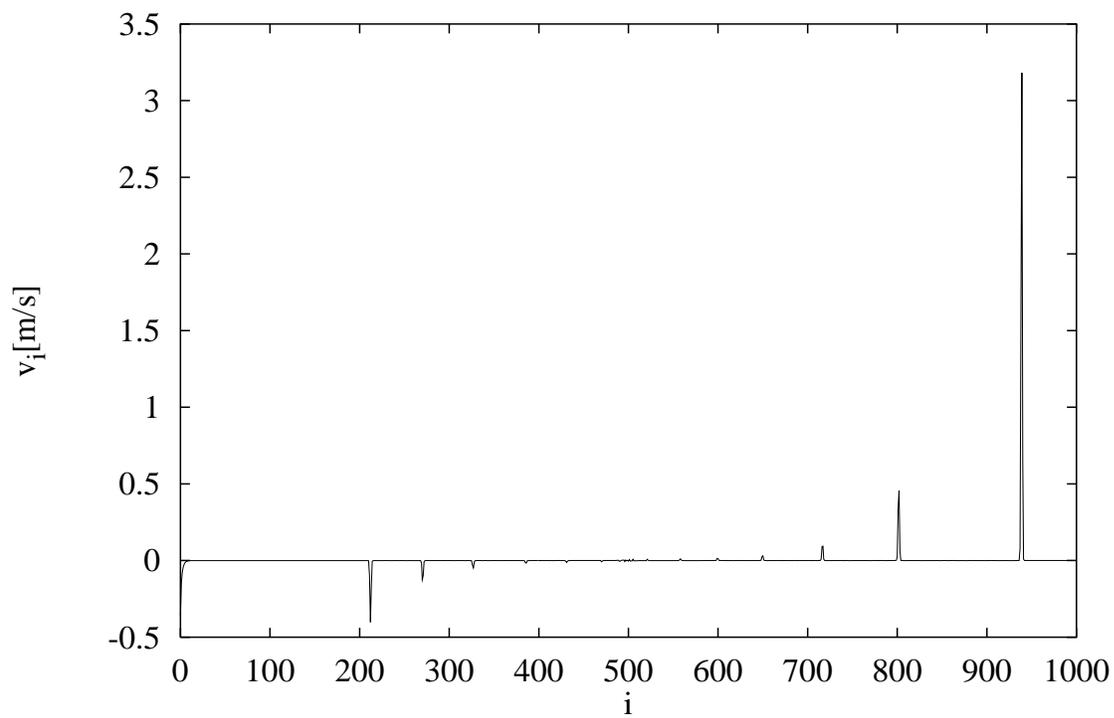,width=\textwidth}}   

\caption[]{Velocities of the $1000$ beads after the scattering of the
shock with a light impurity of mass $m_{i}=0.6g$, the normal mass
being $m=1g$. The force interaction exponent is $\alpha=1.5$. Solitons
are emitted at the impurity location in forward and backward directions.}
\label{vlight}
\end{figure}

\begin{figure}
\centerline{
         \psfig{figure=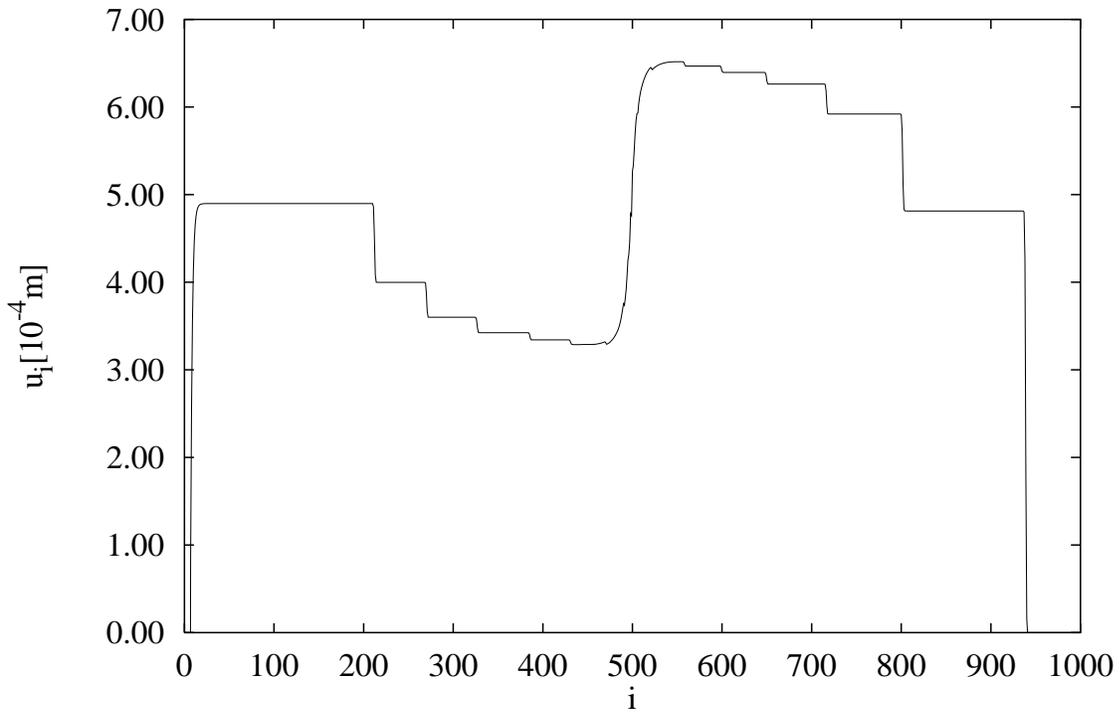,width=\textwidth}}   

\caption[]{Displacement profile corresponding to the velocity profile.}
\label{ulight}
\end{figure}

\begin{figure}[htbp]
\centerline{
         \psfig{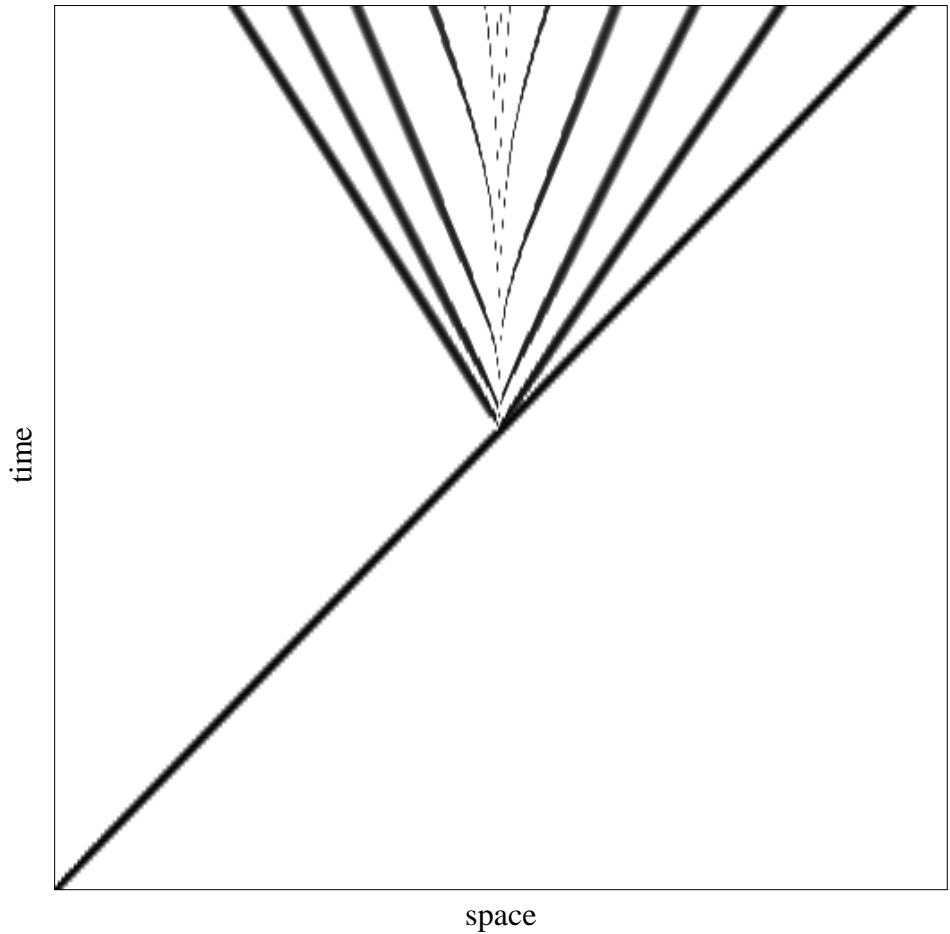}}   

\caption[]{Spatio-temporal evolution of beads contacts for the
collision with a light impurity of mass $m_{i}=0.6g$. A black dot
means that two beads are in compression while a white dot means that
they are at touching. $\alpha$ is equal to $1.5$. Black straight
lines correspond to compression pulses.}
\label{spatio1}
\end{figure}

\begin{figure}[htbp]
\centerline{
         \psfig{figure=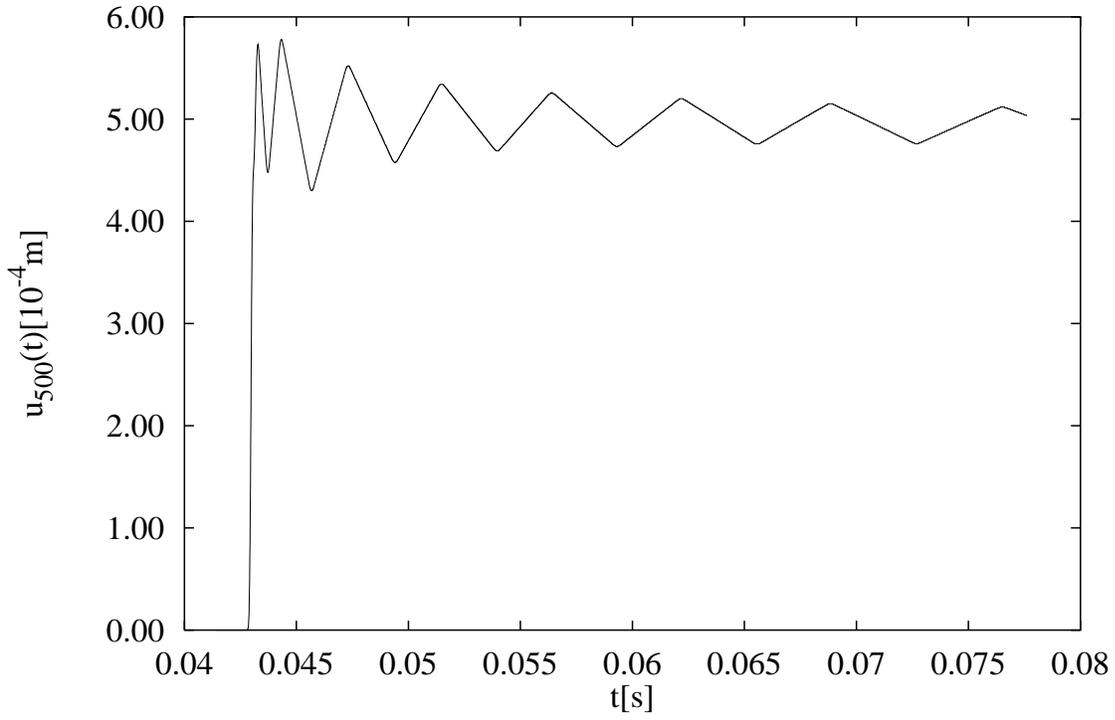,width=\textwidth}}   

\caption[]{Displacement of the impurity located at the position $500$
in function of time for $m_{i}=0.6g$ and $\alpha=1.5$. The light
impurity oscillates in a damped ballistic motion showed by the zig-zag
shape of the curve.}
\label{uimp}
\end{figure}

After having considered the propagation of a single solitary wave in a
chain of equal masses we want to present what are the consequences of
the introduction of a single impurity on the stability of a shock
wave. For that, we consider a chain of $1000$ beads, the bead $500$
being replaced by an impurity. By impurity we mean a bead whose mass
is larger or smaller than the mass of its neighbors. Different
behavior arises as we introduce a light or an heavy impurity. Besides,
we also have to take into account the value $\alpha$ of the
interaction law exponent. The simplest way of looking at what happens
when a shock generated on the left end of the chain scatters the
impurity is to compute the velocity and the displacement of the 
beads after the collision. Fig.(\ref{vlight}) displays velocities for
a light impurity and an exponent $\alpha$ equal to $1.5$. The leading
wave corresponds to the shock wave which is followed by an increasing
number of secondary solitary waves with decreasing amplitudes
generated at the impurity position. A symmetric pattern arises in the
opposite direction since solitary waves with negative velocities and
almost same amplitudes as the previous ones propagate. The wave speed
of the incident shock is modified after the collision since its
amplitude decreases providing with energy the impurity which can then
generate new waves. The corresponding displacement profile is displayed on
Fig.(\ref{ulight}) where the pulses are now replaced by the squared
front waves characteristics of shock waves. In order to better
understand the phenomenon we looked at the spatio-temporal pattern
exhibited by the scattering. In Fig.(\ref{spatio1}) one can see black
straight lines defining compression waves whereas white zones mean
beads are at touching. The secondary pulses are created at the
impurity position and after a transient displayed by thin curved black
lines near the impurity location they propagate with a constant
velocity reaching then a stationary regime. The wave speed of these
pulses are decreasing as new pulses are created as one can see by
looking at the slopes of the black lines. It agrees with the display
of waves with decreasing amplitudes in Fig.(\ref{vlight}). From these
observations we can conclude that the light impurity acts as a wave
source. In order to understand the behavior of this source we computed
the displacement of the impurity as function of time after the
scattering. Fig.(\ref{uimp}) clearly shows that the impurity
oscillates around an equilibrium position. These oscillations are
zig-zag shaped showing that the impurity moves with a constant velocity
between two shocks with its left and right neighbors. This velocity
decreases in time (lower slopes of the oscillations) since the impurity
transfers energy to its neighbors at each shock creating new
pulses. Although we chose special values for the
impurity mass and the interaction law exponent the same behavior is
observed if we maintain the impurity mass lower than the chain masses
and the interaction exponent greater than one.

\begin{figure}[htbp]
\centerline{
         \psfig{figure=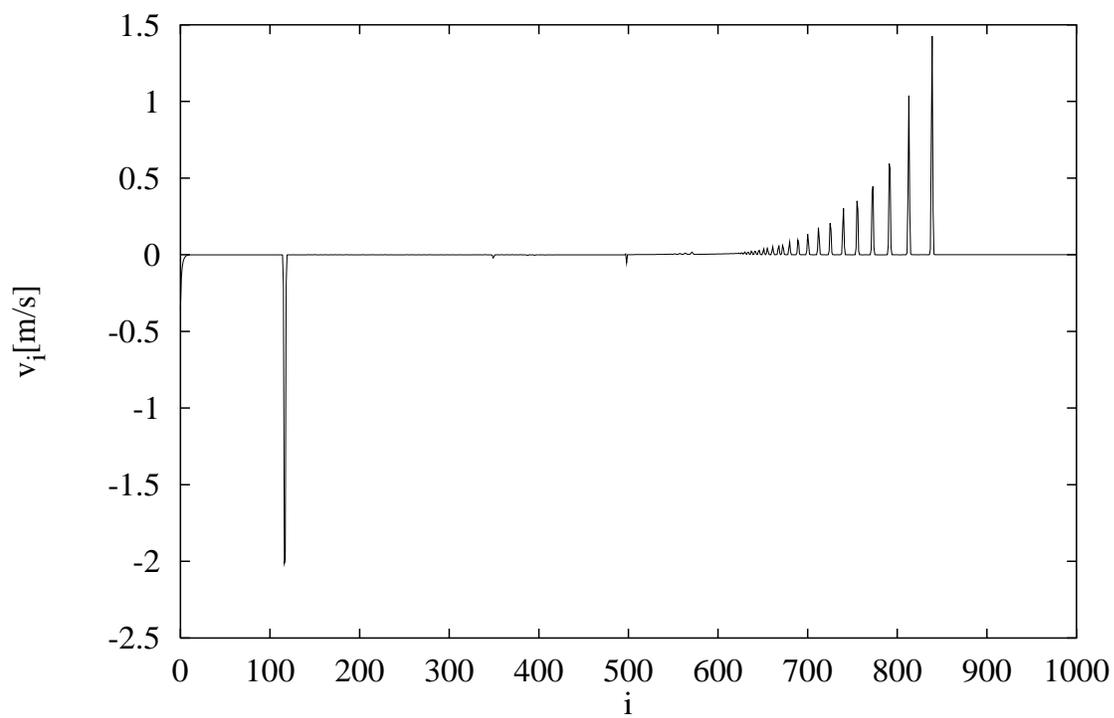,width=\textwidth}}   

\caption[]{Velocities of the $1000$ beads after the scattering of the
shock with a heavy impurity of mass $m_{i}=10g$ ($\alpha=1.5$). The
transmitted solitary wave fragments into many solitary waves of
decreasing amplitudes. The major reflected part is displayed by a
single pulse.}
\label{vheavy}
\end{figure}

\begin{figure}[htbp]
\centerline{
         \psfig{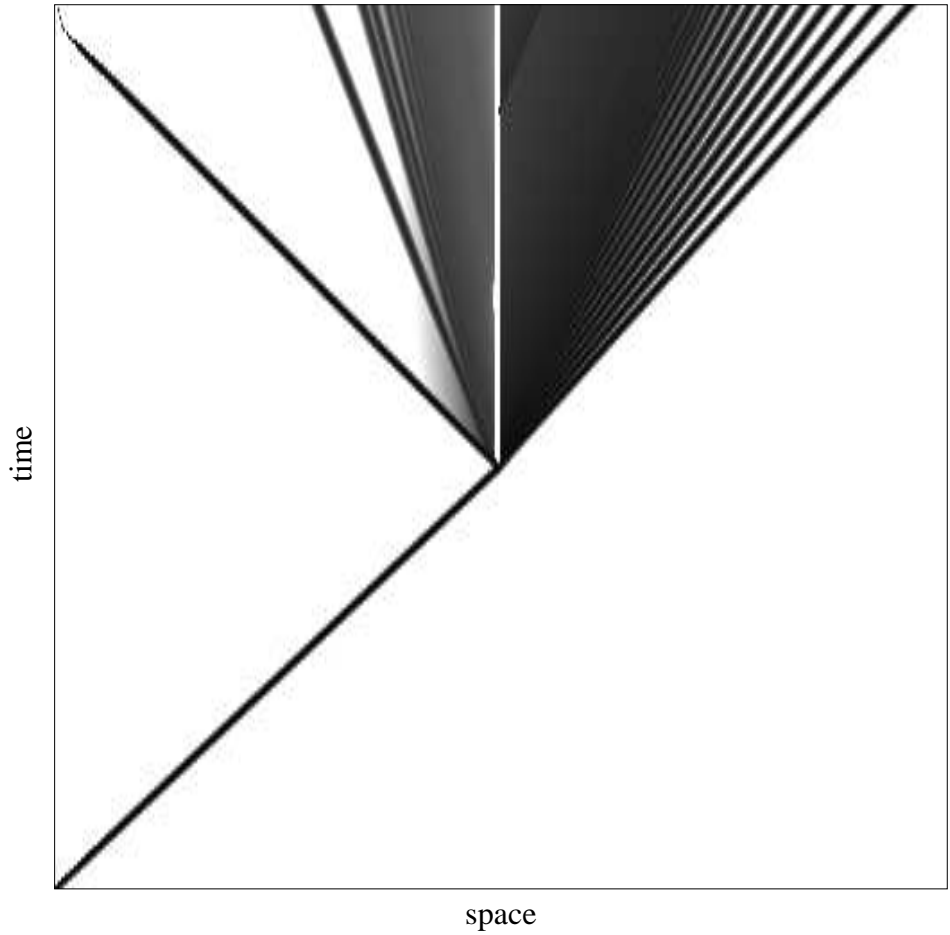}}   

\caption[]{Spatio-temporal evolution of beads contacts for the
collision with an heavy impurity of mass $m_{i}=10g$
($\alpha=1.5$). In addition to the transmitted fragmented front on can
see a second fragmenting front following the reflected wave.}
\label{spatio2}
\end{figure}

\begin{figure}[htbp]
\centerline{
         \psfig{figure=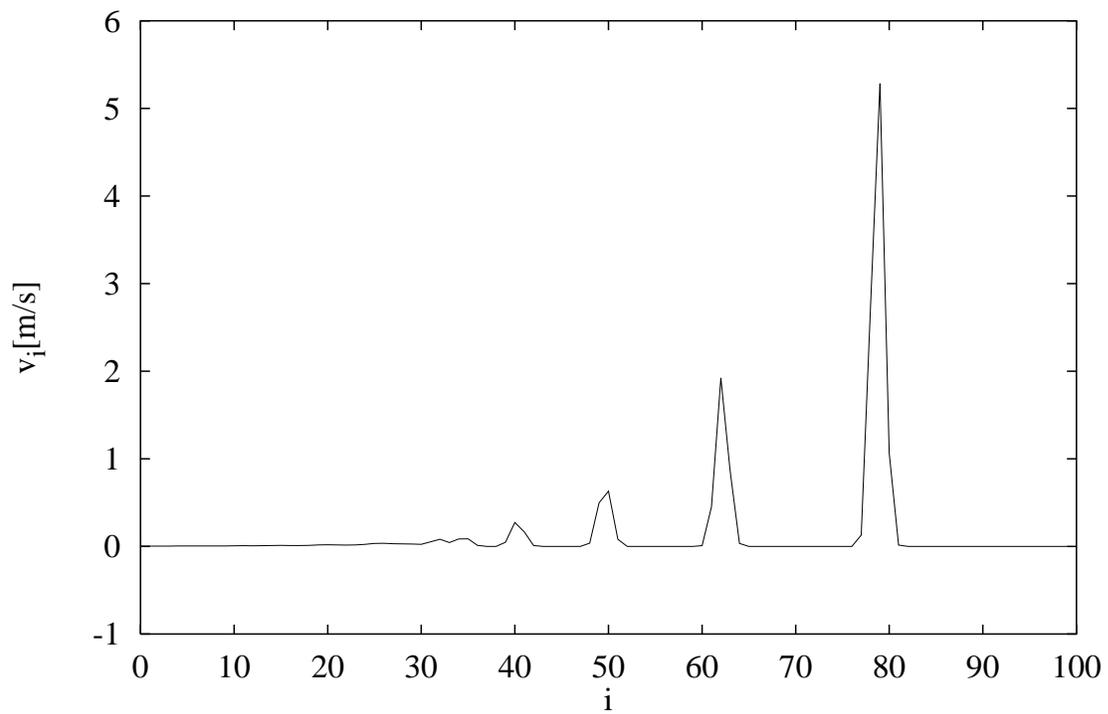,width=\textwidth}}   

\caption[]{Velocities in a $100$ beads chain when it is perturbed with
an heavy mass of $3g$ ($\alpha=1.5$) with initial velocity of
$5m.s^{-1}$. Instead of having a single soliton as it is the case for a
homogeneous chain many pulses of decreasing amplitudes are propagating.}
\label{exp1}
\end{figure}

\begin{figure}[htbp]
\centerline{
         \psfig{figure=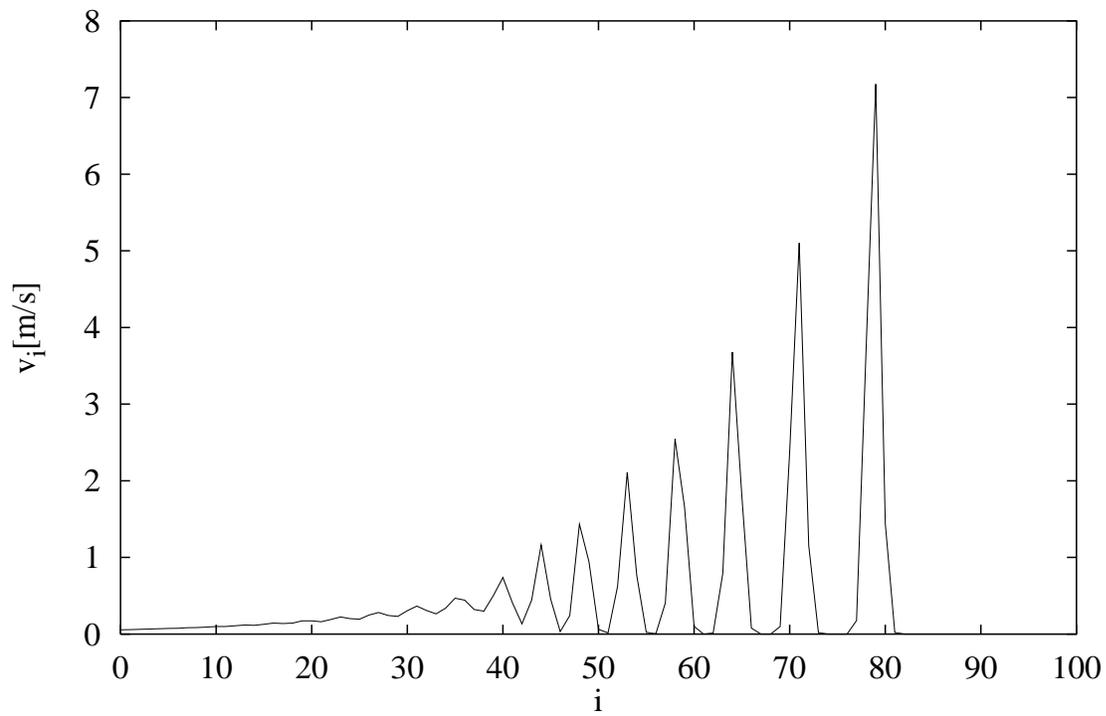,width=\textwidth}}   

\caption[]{Velocities in a $100$ beads chain when it is perturbed with
an heavy mass of $10g$ ($\alpha=1.5$) with initial velocity of
$5m.s^{-1}$. As the mass of the first bead is increased the number of
emitted pulses grows.}
\label{exp2}
\end{figure}

\begin{figure}[htbp]
\centerline{
         \psfig{figure=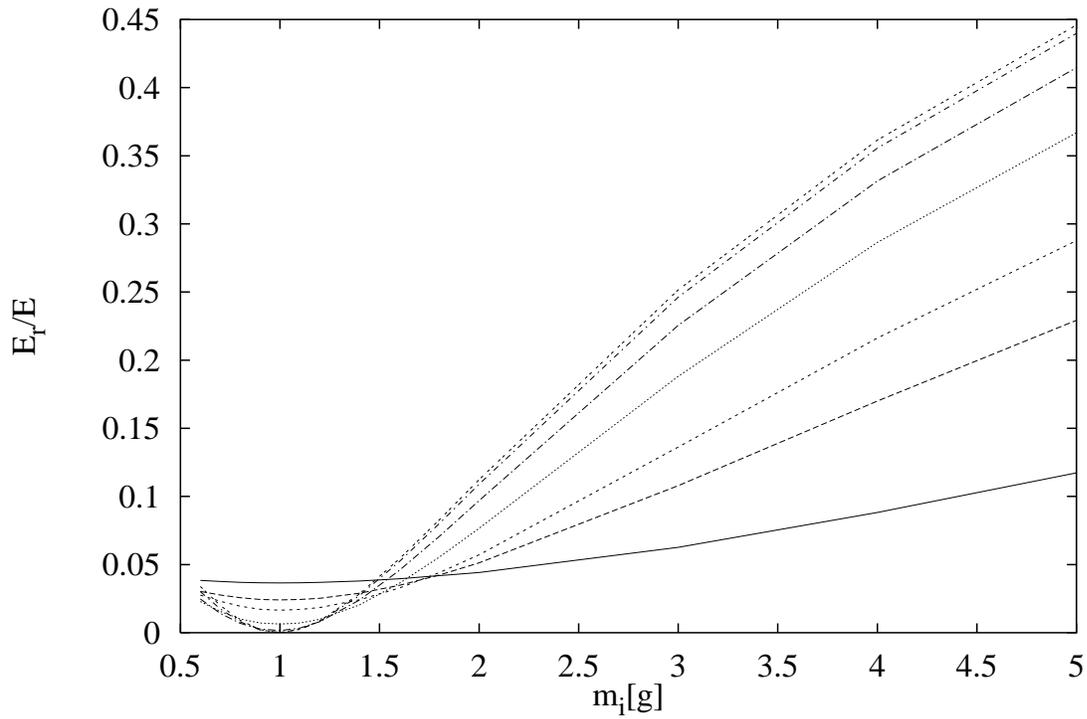,width=\textwidth}}   

\caption[]{Ratios of the reflected energy over
the total energy of the bead chain for different values of
$\alpha$. Going from the upper right curve to the lower right curve
$\alpha$ takes the values $5$, $3$, $2$, $1.5$, $1.2$, $1.1$ and
$1.01$. When $m_i$ is greater than a value between $1.5$ and $2$ the energy
ratio increases with $\alpha$. Below this threshold we have a more
complicated pattern in which curves cross each others at different locations.}
\label{er.e}
\end{figure}

\begin{figure}
\centerline{
         \psfig{figure=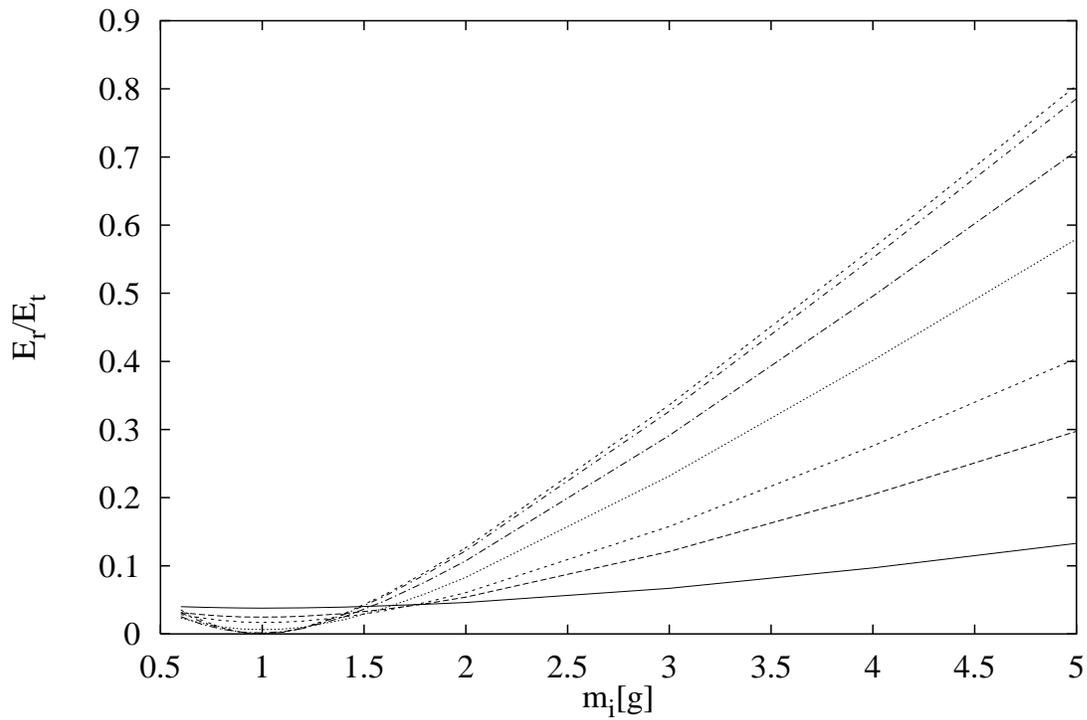,width=\textwidth}}   

\caption[]{Ratios of the reflected energy over the transmitted
energy. Going from the upper right curve to the lower right curve
$\alpha$ takes the values $5$, $3$, $2$, $1.5$, $1.2$, $1.1$ and
$1.01$. A similar pattern as the one for the reflected energy over the
total energy is obtained. The part of reflected energy remains lower
than the part of transmitted energy in the impurity mass considered.}
\label{er.et}
\end{figure}

If we now turn to the case of an heavy impurity we observe a very
different physical picture. In Fig.(\ref{vheavy}) we displayed the
velocity of the beads for an impurity mass greater than beads chain
mass, the interaction exponent remaining equal to $1.5$. The collision
generates many pulses of decreasing amplitudes behind the impurity,
these pulses being very close to each other. Moving in the opposite
direction one can see one pulse of large amplitude. This pattern is
not symmetric as in the case of a light bead and a different physical
process seems to be involved. We can better understand what happened
by looking at the spatio-temporal pattern of contacts displayed in
Fig.(\ref{spatio2}). After collision the bead chain is halved as it is
shown by the white vertical line separating two black areas meaning
compression between beads. The impurity has actually been given a
shift to the right when it collided with the shock wave. Instead of
bouncing back the impurity sticks and there is the formation of a
compression chain. The beads following the impurity don't remain
homogeneously compressed and there is a fragmentation of the incident
pulse giving raise to many pulses corresponding to the straight lines
of decreasing slopes on the spatio-temporal pattern. These lines
obviously correspond to pulses of decreasing amplitude in
Fig.(\ref{vheavy}). Between these pulses the beads remain at rest just
touching their nearest neighbor. Other compression areas appear on the
left side. The straight line of larger slope is the main reflected
wave. A growing compression zone moving at constant velocity follows
this pulse. After some times this area also begins to fragment into
new pulses as one can see from the black straight lines appearing at
the front. These new compression pulses are not visible on the
velocity profile since their amplitude is too low. This
spatio-temporal pattern allows us to say that the collision of a shock
wave with an heavy impurity make it fragment into solitary waves of
decreasing amplitudes. This is not what happens when we introduce a
light impurity since the shock excites the impurity. To our knowledge,
the scattering with an heavy mass is the simplest way to make the
solitary wave unstable. As in the previous case, although we took $10g$ for
the impurity mass and $1.5$ for the interaction exponent the obtained
pattern remains unchanged as these two parameters are tuned. \\
The number of fragmented pulses actually increases with the mass of
the impurity. To show it we considered a chain of equal masses with a
bead of larger mass at its beginning. In the case of a bead having the
same mass we know that the perturbation of the first bead by giving it
a finite velocity at zero time will produce a single pulse. As we
increase the mass of this first bead an increasing number of pulses
appear by the same process as described previously. We plotted the
corresponding velocity profiles for two different values of the first
bead mass in Fig.(\ref{exp1}) and Fig.(\ref{exp2}).  \\
We also looked at the energy distribution in the bead chain as
function of the mass of the impurity and of the value of the force
interaction exponent. Fig.(\ref{er.e}) shows the ratio of the
reflected energy at the impurity location over the total energy in the
chain. The reflected energy is defined as the energy contained in the
first half of the bead chain after scattering. This energy is not
stationary immediately after wave reflection and we have to wait the
time for it to reach a stationary value. The
impurity mass varies between $0.6g$ and $5g$ and $\alpha$ takes values
between $1.01$ and $5$. One easily sees that the amount of reflected
energy increases as the mass difference increases. This amount also
increases with $\alpha$ if the the impurity mass is greater than a
value located between $1.5g$ and $2g$. Below this value the curves
cross each others at different values of the impurity mass. Below the
crossing it is the lowest value of the two exponents which gives the
greater amount of reflected energy. If we now consider the ratio of
reflected energy over transmitted energy in Fig.(\ref{er.et}) we
observe similar variations as previously. Moreover the ratios remain
lower than unity meaning that the amount of reflected energy is always
lower than the amount of transmitted energy in the impurity mass
interval considered.

\section{Conclusion}

We investigated the problem of a shock propagating in a non-loaded
horizontal granular chain with a general beads interaction force $F
\propto \delta^\alpha$, $\alpha>1$. We compared our numerical
simulations with previous analytical results obtained in the long
wavelength approximation \cite{Nest2}. For values of $\alpha$ close to
unity we found a cross-over between a non-linearity dominated regime
and a solitonic one corresponding to the long wavelength
approximation. The latter regime is reached after a time which
increases and tends to infinity as one tunes $\alpha$ to unity. It was
then shown that the transition between the $\alpha=1$ \cite{HHL,HS}
(self-similar) regime to the $\alpha>1$ regime is continuous. In the
case of large values of $\alpha$ we observed discrepancies between
numerical simulations and analytical results. As $\alpha$ increases
beyond $1.2$ tails in the deformation curves separate. When $\alpha$
reaches the value of $1.7$ the analytical result given for the
deformation curve is already not reliable. In the limit of very large
values of $\alpha$ the deformation curve takes a triangular shape
totally inconsistent with the one given by the theoretical analysis.\\
We then studied the interaction of these solitonic shock waves with an
impurity mass placed in the middle of the bead chain. We first
considered light impurities. The scattering of the shock with the
impurity actually excites it in an oscillatory ballistic damped
motion. The impurity acts as a wave source emitting solitary waves of
decreasing amplitude in both directions. If we place an heavy impurity
in the bead chain we observe a completely different pattern. The
impurity is no more excited but sticks its next neighbor splitting the
bead chain in two parts. The chain corresponding to the transmitted
wave supports no more solitonic propagation and we have an instability
displayed by the fragmentation of the transmitted wave into waves of
decreasing amplitudes. On the other half of the chain a reflected
stable solitonic wave propagates followed by a growing unstable
fragmented front appearing at the impurity location. The physics
involved in this case is much more complicated and deserves further
investigations.

\newpage

\end{document}